\documentclass[aps,prl,reprint,superscriptaddress]{revtex4-2}

\usepackage{xcolor}
\usepackage{graphicx}
\usepackage{dcolumn}
\usepackage{bm}
\usepackage{amsmath}

\begin{document}


\title{Symmetry Breaking and Restoration in Turbulent Thermal Convection Arises from the Competition Between Advection and Buoyancy}

\author{Guang-Yu Ding}
 \email{dinggy@fudan.edu.cn}
 \affiliation{Center for Complex Flows and Soft Matter Research and Department of Mechanics and Aerospace Engineering, Southern University of Science and Technology, Shenzhen, China}%
 \affiliation{Research Institute of Intelligent Complex Systems, Fudan University, Shanghai, China}%

\author{Fang Xu}%
\affiliation{Center for Complex Flows and Soft Matter Research and Department of Mechanics and Aerospace Engineering, Southern University of Science and Technology, Shenzhen, China}%

\author{Ke-Qing Xia}%
\email{xiakq@sustech.edu.cn}
\affiliation{Center for Complex Flows and Soft Matter Research and Department of Mechanics and Aerospace Engineering, Southern University of Science and Technology, Shenzhen, China}%
\affiliation{Department of Physics, Southern University of Science and Technology, Shenzhen, China}%

\date{\today}

\begin{abstract}
Spontaneous symmetry breaking (SSB) remains poorly understood in thermal convection, but hints may be found from its restoration. We hereby compare the two convection systems: experiments with polymer additives, and simulations with linear friction. We observe the restoration of similar symmetric flows in both these systems. Additionally, restoration coincides with enhanced, time-symmetric velocity-buoyancy correlation, and a sharp drop in the normalized buoyancy-response time. These results indicate buoyancy predominance: velocity is statistically slaved to buoyancy and preferentially remains vertical. The predominance of buoyancy provides a local orientation mechanism, which is necessary for restoring the symmetry of the system. Conversely, this orientation mechanism is lost locally in canonical convective flows, thus SSB naturally occurs in Rayleigh-B\'{e}nard convection. Our results suggest that the breaking and restoration of symmetry in thermal convection are both attributable to the competition between advection and buoyancy.

\end{abstract}

\maketitle

Symmetry plays an essential role in physics, most famously through Noether's theorem \citep{Noether1918}, which links continuous symmetries to conservation laws. Spontaneous symmetry breaking (SSB) occurs even when the governing equations and boundary conditions remain symmetric \citep{Gross1996, Cantwell2002, Schwichtenberg2018, Williams2025}. Turbulent Rayleigh--B\'{e}nard convection (RBC), a paradigm for thermally driven turbulence \citep{Ahlers2009b, Lohse2010, Chilla2012, Xia2013}, provides a clear example: in a cylindrical cell, the equations and boundary conditions are invariant under rotations about the vertical axis, yet the observed large-scale flow is typically not axisymmetric, as first revealed by \citet{Krishnamurti1981}. Instead, thermal plumes emitted from the boundary layers self-organize into a single-roll structure, namely the large-scale circulation (LSC), thereby breaking the rotational symmetry \citep{Sano1989, Gluckman1993, Zhang1997a, Verzicco1999, Xi2004}. Despite extensive studies, the physical origin of this SSB in RBC remains poorly understood \citep{Kadanoff2001}. In particular, it is unclear which mechanism selects a non-axisymmetric LSC over an axisymmetric state allowed by the symmetries of the problem.

Clues may come from the converse phenomenon: flow symmetry restoration. Recently, \citet{xu_2025_prl} reported that minute polymer additives can restore an axisymmetric mean flow topology in turbulent RBC. This restored state is accompanied by anisotropic suppression of turbulent fluctuations: the root-mean-square (rms) horizontal velocity decreases more strongly than the vertical component. A key open question is whether this anisotropic suppression causes symmetry restoration or is instead a consequence of a deeper dynamical process. Because polymer-laden turbulence is complex and remains far from fully understood \citep{Peng1983, Tabor1986}, this question is difficult to address directly with viscoelastic models. Although various numerical models of viscoelastic turbulence have been proposed \citep{Herrchen1997, Benzi2018, alves_arfm_2021_nmvff, Serafini2022, Datta2022, song_jcp_2025_fdmttc}, they do not provide a simple way to independently control the observed suppression. We therefore introduce a minimal model in which a two-dimensional RBC incorporates linear friction that can be imposed either isotropically or anisotropically. This lets us test directly whether anisotropic suppression alone is sufficient to recover a symmetric flow state. Although linear friction acts at large scales, whereas polymer additives are generally considered to act at small scales \citep{Bird1987a, Perkins1995, Benzi2010, benzi_arcmp_2018_pff, zhang_pnas_2025_eecteea}, the two systems show unexpected similarities.

The experiments use a cylindrical convection cell and long-chain polyacrylamides similar to those in \citet{Xu2024, xu_2025_prl, Xu2026}. The Rayleigh number $Ra\equiv\beta_T g\Delta H^3/(\kappa\nu)$ is fixed at $3\times10^9$, and the Prandtl number $Pr\equiv\nu/\kappa$ is 4.34. Here $\Delta$ is the temperature difference across the fluid layer, $H$ the cell height, $g$ gravity, and $\beta_T$, $\kappa$, and $\nu$ are the thermal expansion coefficient, thermal diffusivity, and kinematic viscosity, respectively. The polymer concentration $c$ varies from 0 to 20~ppm (parts per million by weight). For the simulations, we use direct numerical simulation (DNS) to study two-dimensional RBC with linear friction in two configurations: isotropic friction (IF, $\alpha_x=\alpha_z=\alpha$) and anisotropic friction (AF, $\alpha_x=\alpha$, $\alpha_z=0$). The linear friction term $\mathcal{F}$ is given by
\begin{equation}
	\mathcal{F}\equiv (\alpha_x u_x\hat{e}_x+\alpha_z u_z\hat{e}_z)/\tau_{ff}.
\end{equation}
Here $u_x$ and $u_z$ are horizontal and vertical velocity components, $\hat{e}_x$ and $\hat{e}_z$ are the corresponding unit vectors, and $\alpha_x$ and $\alpha_z$ are dimensionless friction coefficients. The reference coefficient is $\alpha=\tau_{ff}/\tau_\alpha$, where $\tau_{ff}=\sqrt{H/(\beta_T g\Delta)}$ is the free-fall time scale and $\tau_\alpha=1/\alpha^*$ is the frictional time scale associated with the dimensional coefficient $\alpha^*$. We consider $\alpha$ from $1\times10^{-2}$ to $4\times10^1$, $Ra$ from $10^8$ to $10^{10}$, and $Pr=1$. The sidewalls are adiabatic and all boundaries are no-slip. The governing equations are solved using the well-tested CUPS code, based on a fourth-order finite-volume method \citep{chong_jcp_2018}. For simplicity, velocity $u$ and temperature $T$ denote dimensionless quantities normalized by $u_{ff}$ and $\Delta$.

\begin{figure}
	\centerline{\includegraphics[width=\columnwidth]{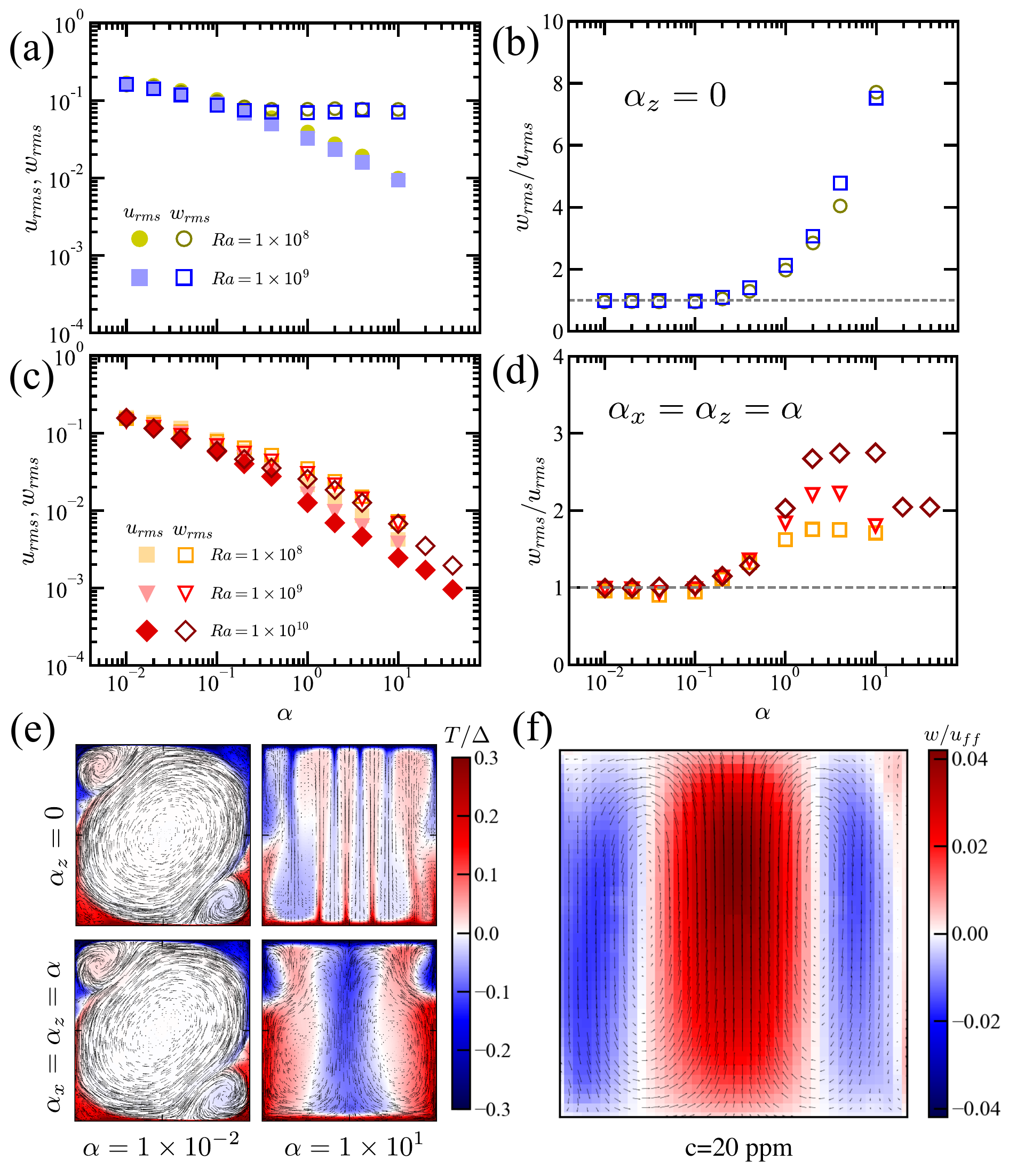}}
	\caption{(a,c) Root-mean-square (rms) vertical velocity $w_{rms}$ and horizontal velocity $u_{rms}$. Closed circles and open symbols denote $u_{rms}$ and $w_{rms}$, respectively. (b,d) The rms ratio $w_{rms}/u_{rms}$. Upper panel corresponds to anisotropic friction ($\alpha_x=\alpha$, $\alpha_z=0$), while middle panel corresponds to isotropic friction ($\alpha_x=\alpha_z=\alpha$). Mean temperature and velocity vectors from the numerical simulation (e), and normalized vertical velocity from the polymer-laden experiment with $c=20$ ppm (f). Here $u_{ff}$ denotes the free-fall velocity.}
	\label{fig:flow_structure}
\end{figure}

We first clarify the relationship between anisotropic suppression and flow symmetry. Figures \ref{fig:flow_structure}(a,c) show the rms horizontal and vertical velocities, $u_{rms}$ and $w_{rms}$, for anisotropic friction (AF, $\alpha_z=0$) and isotropic friction (IF, $\alpha_x=\alpha_z=\alpha$), respectively. Figs.~\ref{fig:flow_structure}(b,d) show the ratio $w_{rms}/u_{rms}$. In both AF and IF, weak friction reduces velocity rms while the flow remains LSC-like in Fig.~\ref{fig:flow_structure}(e), $\alpha=1\times10^{-2}$. As $\alpha$ increases beyond $\sim 10^{-1}$, $w_{rms}/u_{rms}$ rises monotonically in AF. In IF it rises initially, indicating the emerging anisotropic suppression of horizontal velocity, then saturates at $w_{rms}/u_{rms}\sim 2$. Eventually, a symmetric flow is restored, as shown in the lower-right panel of Fig.~\ref{fig:flow_structure}(e), which closely resembles the polymer experiment at $c=20$~ppm in Fig.~\ref{fig:flow_structure}(f). Although increasing $w_{rms}/u_{rms}$ coincides with more symmetric patterns, full symmetry restoration is observed only in IF. AF yields a much larger $w_{rms}/u_{rms}$ but produces less symmetric multi-column structures. Therefore, anisotropic suppression appears to be a consequence of symmetry restoration, not its cause. Additionally, frictionless simulations at reduced $Ra$ show that simply weakening turbulence does not recover symmetry, implying a deeper dynamical process driving both anisotropic suppression and symmetry restoration.

\begin{figure}
\centerline{\includegraphics[width=\columnwidth]{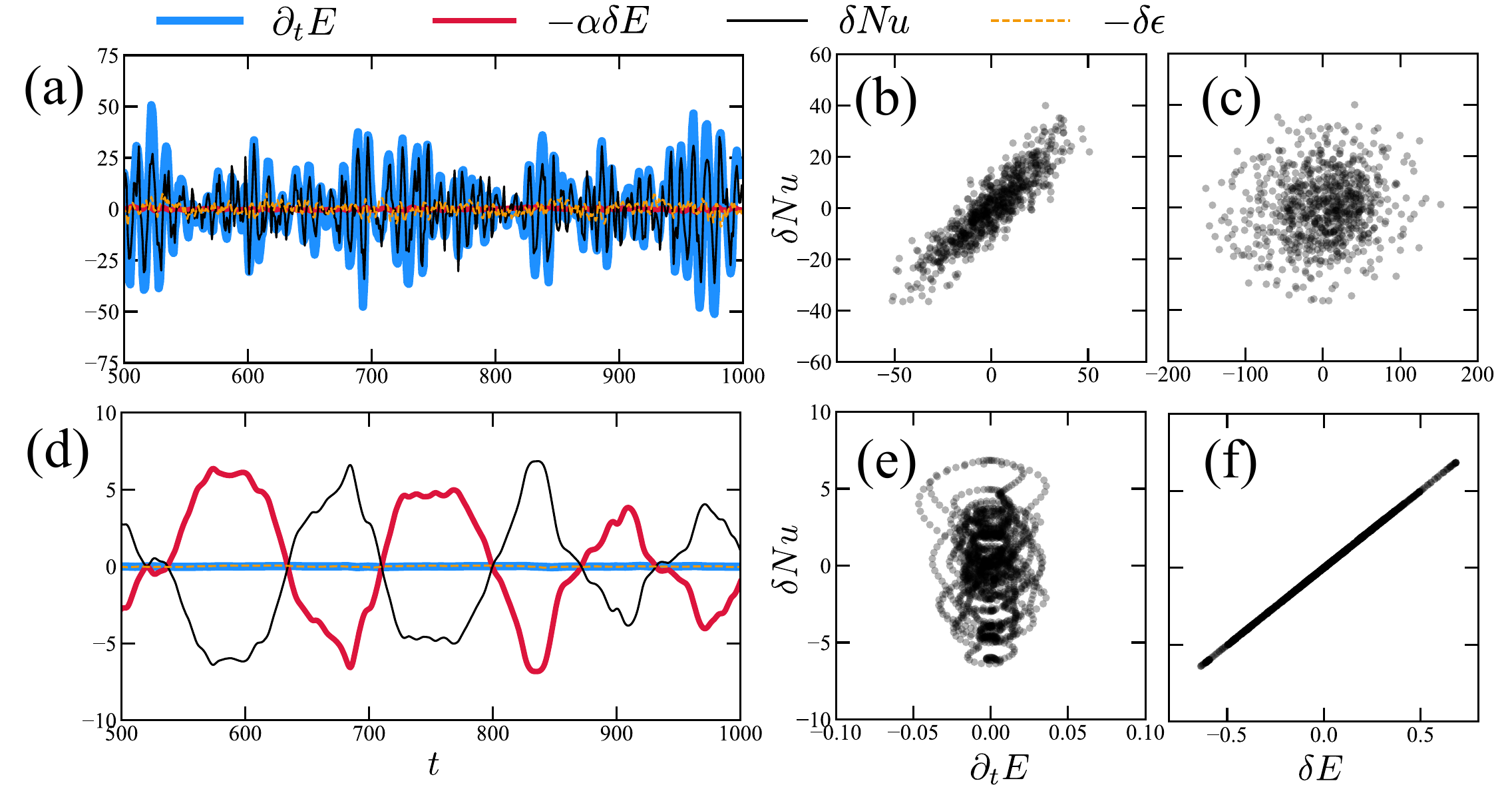}}
\caption{Time traces of the terms in the dynamical balance relation \ref{equ:dynamical_balance} (a,d), scatter plots of $\partial_t E$ versus $\delta Nu$ (b,e), and scatter plots of $\delta E$ versus $\delta Nu$ (c,f), for (a-c) $\alpha=10^{-2}$ and (d-f) $\alpha=10^{1}$. Here $E$ is the global kinetic energy and $Nu$ is the Nusselt number.}
\label{fig:dyn_balance}
\end{figure}

As the numerical model is much easier to interpret, we first focus on how symmetry is restored in the frictional case, and return to the polymer system later. Let $E= \langle u_i^2\rangle$ be the global kinetic energy, $\epsilon= Pr \langle(\partial_j u_i)^2\rangle$ the dimensionless viscous dissipation rate, and $Nu=\sqrt{RaPr}\langle uT\rangle+1$ the Nusselt number. Taking the global average of the energy equation and subtracting the mean components gives
\begin{equation}
    \label{equ:dynamical_balance}
    \frac{\partial E}{\partial t}=-\delta\epsilon+\delta Nu-\alpha \delta E,
\end{equation}
where $\delta\epsilon\equiv\epsilon-\overline{\epsilon}$ is the fluctuation around the temporal mean, and $\delta Nu$ and $\delta E$ are defined analogously. The time traces in Figs.~\ref{fig:dyn_balance}(a,d) show two distinct regimes. For $\alpha=10^{-2}$ (LSC state), $\partial_tE$ is dynamically similar to $\delta Nu$ with comparable magnitude, and the scatter plot in Fig.~\ref{fig:dyn_balance}(b) confirms that global kinetic-energy variations are mainly attributed to buoyancy input. For $\alpha=10^{1}$, corresponding to the symmetric case in Fig.~\ref{fig:flow_structure}(e), $\partial_tE$ is weak because of strong frictional damping. The buoyancy input is then balanced by the frictional dissipation $\alpha \delta E$, as illustrated by the scatter plot between $\delta E$ and $\delta Nu$ in Fig.~\ref{fig:dyn_balance}(f).

\begin{figure*}
	\centerline{\includegraphics[width=\textwidth]{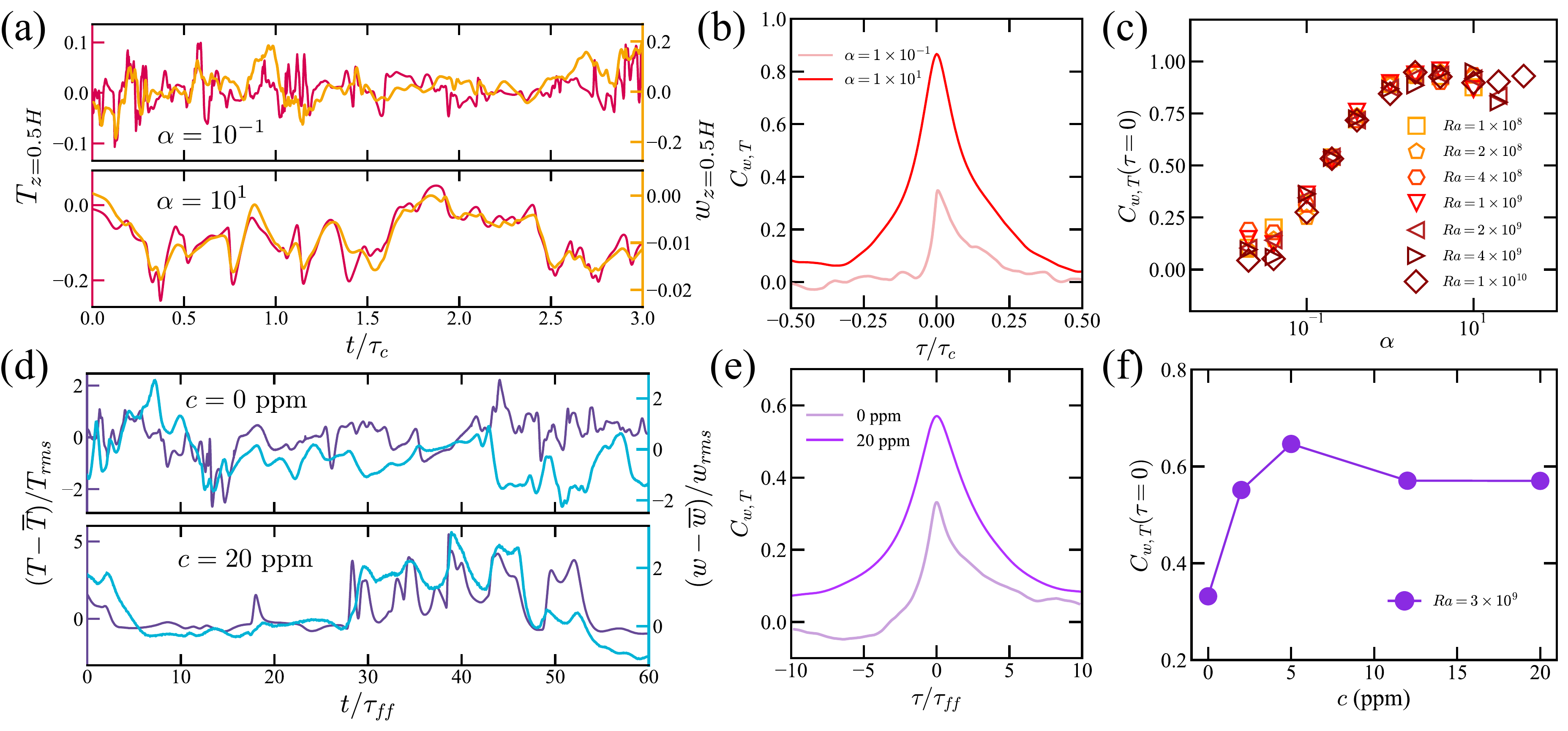}}
	\caption{Local temperature and velocity data for the simulation with linear friction (upper panels) and the experiment with polymer additives (lower panel). (a,d) Time traces of vertical velocity $w$ and temperature $T$ at the cell center, (b,e) the local cross-correlation function between vertical velocity and temperature, $C_{w,T}$, and (c,f) the cross-correlation coefficient $C_{w,T}(\tau=0)$ as a function of $\alpha$ for the RBC system with friction and polymer concentration $c$ for the experiments, respectively. The convective time scale $\tau_c$ in (a,b) is defined as $\tau_c\equiv(\int \hat{w}^2/k dk)/(\int\hat{w}^2 dk)$, where $\hat{w}$ is the Fourier transform of the time series of $w$ and $k$ is the wavenumber.}
	\label{fig:uT_two_sys}
\end{figure*}

The change in the dynamical energy balance implies an evolution from indirect velocity-buoyancy coupling to the direct one. Dynamically, the weak-friction regime gives
\begin{equation}
    \label{equ:indirect_coupling}
    uT\sim\frac{\partial u^2}{\partial t}\sim u\frac{\partial u}{\partial t},\text{\quad or\quad} T\sim\partial_t u,
\end{equation}
whereas the strong-friction regime approaches direct coupling,
\begin{equation}
    \label{equ:direct_coupling}
    uT\sim u^2,\text{\quad or\quad}T\sim u.
\end{equation}

In the former case, buoyancy introduces a vertical accelaration, but buoyancy and velocity are not necessarily aligned. In the direct-coupling case, buoyancy and velocity are directly correlated, thus velocity vectors are preferentially vertical. This explains why isotropic friction still yields an anisotropic suppression ($w_{rms}/u_{rms}>1$). Together with the suppression of small-scale velocity fluctuations, this vertical preference promotes large buoyancy-aligned structures, consistent with the symmetric structures in Figs.~\ref{fig:flow_structure}(e,f).

Unlike in the frictional case, the direct-coupling relation in Eq.~\eqref{equ:direct_coupling} is not mathematically guaranteed in the polymer-laden system. Nevertheless, an axisymmetric structure still requires a local orientation mechanism that distinguishes horizontal and vertical directions: In thermal convection, buoyancy is the natural candidate. We therefore expect the correlation between velocity and buoyancy (or local temperature) to be a unifying feature of both the frictional model and the polymer experiment. Figures~\ref{fig:uT_two_sys}(a) and (d) compare the cell-center time traces of velocity and temperature in the two systems. For $\alpha=10^{-1}$ (LSC state), the signals agree on large scales but deviate at small scales; the same pattern appears for $c=0$~ppm in Fig.~\ref{fig:uT_two_sys}(d). For $\alpha=10^{1}$ (symmetric state), velocity closely follows temperature except for some high-frequency features, i.e., velocity becomes a low-pass filtered version of temperature. Analogous behavior is seen with polymer additives at $c=20$~ppm. The traces are not perfectly aligned but remain roughly in phase, suggesting a change in velocity-buoyancy coupling.

We quantify this change using the cross-correlation function $C_{w,T}=\langle w^\prime(t+\tau)T^\prime(t)\rangle/(w_{rms}T_{rms})$, where primes denote fluctuations and $\tau$ is the lag. Cell-center measurements of $C_{w,T}$ are shown in Figs.~\ref{fig:uT_two_sys}(b) and (e). For $\alpha=10^{-1}$ and $c=0$~ppm, $C_{w,T}$ is asymmetric in $\tau$. It rises sharply as $\tau\to 0^{-}$ and decays more slowly for $\tau>0$. This is consistent with canonical plume dynamics, in which buoyancy driving induces a long-tail velocity response \citep{Shang2004}. For $\alpha=10^{1}$ and $c=20$~ppm, $C_{w,T}(\tau=0)$ increases significantly. Figures~\ref{fig:uT_two_sys}(c) and (f) plot $C_{w,T}(\tau=0)$ against friction strength and polymer concentration, respectively. In both systems, $C_{w,T}(\tau=0)$ rises and reaches a plateau where symmetry restoration is observed, showing that restoration is accompanied by stronger temperature-velocity correlation.

Another important feature is that the cross-correlation becomes nearly time-symmetric for $\alpha=10^1$ and $c=20$ ppm, which is a significant change in the nature of the temperature-velocity relationship. This time-symmetric feature implies that temperature cannot be treated as a simple `active' scalar, since past velocity cannot respond to future temperature, no matter how `active' it is. In the frictional system, this feature follows from the direct coupling in Eq.~\eqref{equ:direct_coupling}. With weak $\partial u/\partial t$, evolution is governed by temperature, so the system is buoyancy-predominant and velocity is slaved to buoyancy. In the polymer-laden system, the same energy balance is not guaranteed. A plausible explanation is that polymer additives shorten the buoyancy-response time $\tau_b$, producing a rapid velocity response to temperature variation. When the response is nearly instantaneous, the cross-correlation between $w$ and $T$ becomes auto-correlation-like and time-symmetric, leading to a similar picture as the frictional case.

\begin{figure}
	\centerline{\includegraphics[width=\columnwidth]{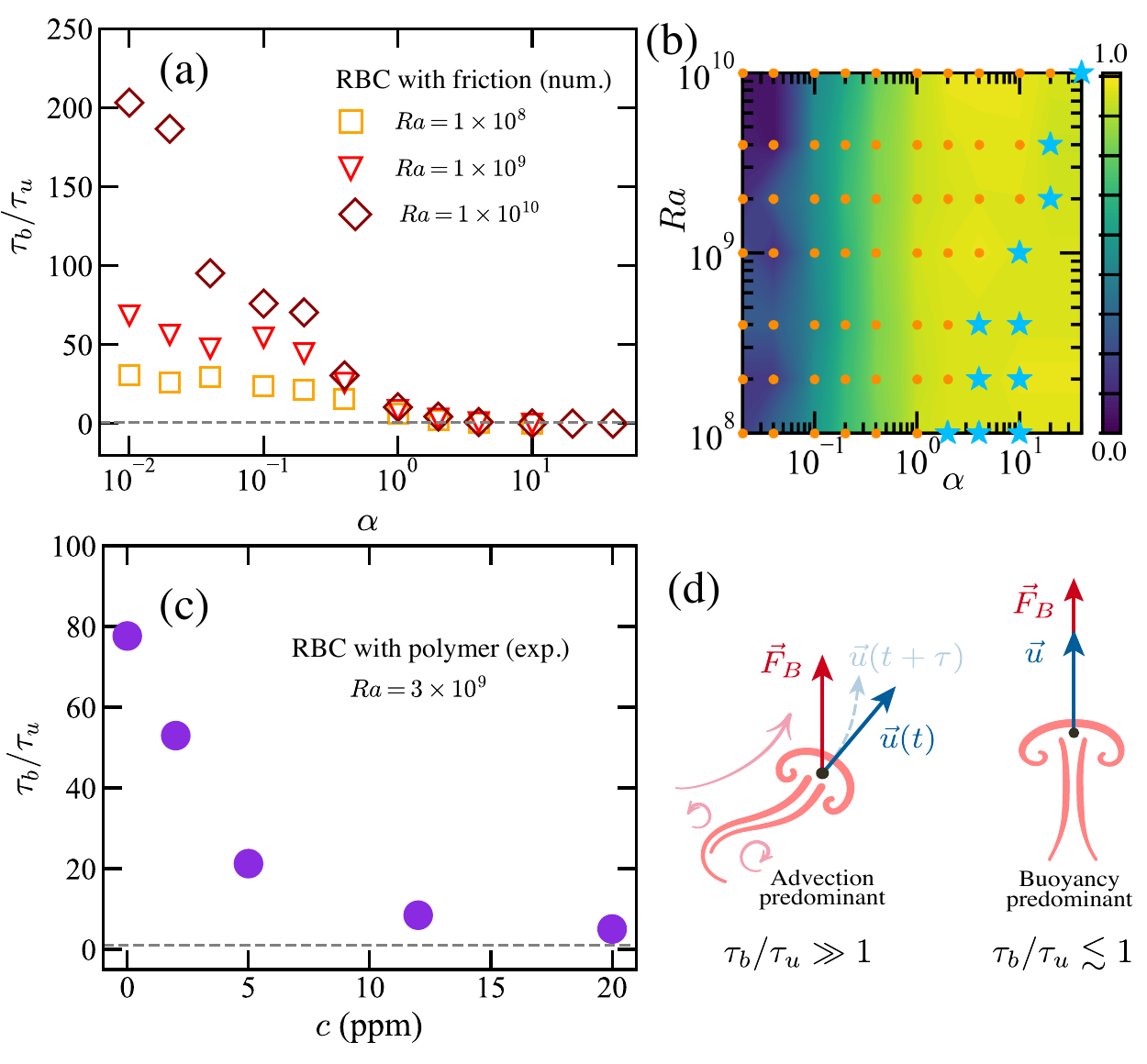}}
	\caption{The ratio of the buoyancy-response time $\tau_b$ to the advection time $\tau_u$ for (a) simulations with linear friction and (c) polymer experiments. The grey dashed lines indicate $\tau_b/\tau_u=1$. In (b) we show $C_{w,T}(\tau=0)$, where orange circles denote the simulations in this study and blue stars indicate cases with symmetric structure. In (d), we illustrate plume behavior in advection- and buoyancy-predominant scenarios, respectively. Here $\vec{F}_B$ denotes the buoyancy vector.}
	\label{fig:time_argument}
\end{figure}

To examine this argument, we consider the following time-scale ratio. For a plume in the bulk, the advection time scale $\tau_u$ and the response time scale to buoyancy variation $\tau_b$ can be written, respectively, as
\begin{equation}
	\tau_u\sim \frac{\Delta l}{w_{rms}},\ \tau_b\sim\frac{w_{rms}}{\beta_T g T_{rms}}.
\end{equation}
The plume length scale is estimated as $\Delta l\approx 0.5H/Nu$ from the thermal boundary-layer thickness. This gives
\begin{equation}
	\label{equ:time_ratio}
	\frac{\tau_b}{\tau_u}\sim 2Nu\left(\frac{w_{rms}}{u_{ff}}\right)^2\left(\frac{\Delta}{T_{rms}}\right).
\end{equation}
When $\tau_b/\tau_u\gg 1$, it corresponds to an advection-predominant picture, in which plumes are advected by the ambient flow, and thus the influence of buoyancy on the velocity vector is limited. When $\tau_b/\tau_u$ is small, the velocity responds rapidly and vertical alignment is strengthened. Figure~\ref{fig:time_argument} shows that $\tau_b/\tau_u$ decreases with increasing $\alpha$ in the frictional system (as expected from direct coupling), and with increasing $c$ in the polymer experiment. Since cell-center measurement may overestimate the influence of advection for the whole system, $\tau_b/\tau_u$ remains a bit above unity at the largest $c$ for the polymer case. Nevertheless, together with the correlation analysis above, the trend supports the transition to the buoyancy-predominant dynamics, consistent with the buoyancy-aligned structures in Figs.~\ref{fig:flow_structure}(e,f).

In conclusion, our results show that flow symmetry restoration in turbulent thermal convection is not simply caused by anisotropic suppression. Instead, the comparison between frictional simulations and polymer-laden experiments suggests buoyancy predominance as the underlying mechanism. In the frictional system, this appears as direct coupling between buoyancy and velocity. In the polymer-laden system, the enhanced temperature-velocity correlation, the nearly time-symmetric cross correlation function, and the strong reduction of $\tau_b/\tau_u$ all suggest a similar statistical picture. Thus, both linear friction and, more importantly, polymer additives promote buoyancy alignment of the flow. This orientation mechanism explains why symmetric structures can form, and why symmetry restoration cannot be achieved merely by weakening turbulence. The microscopic mechanism by which polymer additives induce this rapid response to buoyancy remains unclear. One way could be through the more coherent thermal plumes, as they become more buoyant objects compared to the case without polymers \citep{Benzi2010, Xie2015}.

This picture also suggests a possible interpretation of spontaneous symmetry breaking in thermal convection. If symmetry is restored through buoyancy predominance and the accompanying orienting mechanism, then SSB may be associated with the loss of this local orientation. In fact, previous experiments have shown that temperature exhibits a passive-like behavior in the bulk \citep{Moisy2001, zhou_prl_2002_psttma} due to the advection dominance, as indicated by the large $\tau_b/\tau_u$ ratio at $c=0$ ppm in Fig.~\ref{fig:time_argument}(c). It suggests that the local flow cannot effectively distinguish the vertical direction, even in a weakly turbulent background at small $Ra$. In this sense, the local flow and the global system possess different effective symmetries, which naturally explains how SSB occurs in thermal convection.

This study is supported by the National Natural Science Foundation of China (NSFC) through grants (Nos. 12302282, 12572251, 12594530010, 12202174, 12595302), the Startup Foundation of Fudan University, and the Center for Computational Science and Engineering of Southern University of Science and Technology. G.Y.D. and F.X. contributed equally to this work.

\bibliography{reference}

@article{alves_arfm_2021_nmvff,
  title = {Numerical Methods for Viscoelastic Fluid Flows},
  author = {Alves, M.A. and Oliveira, P.J. and Pinho, F.T.},
  year = 2021,
  month = jan,
  journal = {Annu. Rev. Fluid Mech.},
  volume = {53},
  number = {1},
  pages = {509--541},
  issn = {0066-4189, 1545-4479},
}

@article{benzi_arcmp_2018_pff,
  title = {Polymers in Fluid Flows},
  author = {Benzi, R. and Ching, E. S. C.},
  year = 2018,
  journal = {Annu. Rev. Condens. Matter Phys.},
  volume = {9},
  number = {1},
  pages = {163--181},
  issn = {1947-5454 1947-5462},
  chapter = {163},
}

@article{chong_jcp_2018,
	title = {Multiple-resolution scheme in finite-volume code for active or passive scalar turbulence},
	author = {Chong, K. L. and Ding, G. and Xia, K.-Q.},
	year = {2018},
	journal = {J. Comput. Phys.},
	volume = {375},
	pages = {1045--1058}
}

@article{song_jcp_2025_fdmttc,
  title = {A Finite Difference Method for Turbulent Thermal Convection of Complex Fluids},
  author = {Song, J. and Xu, C. and Shishkina, O.},
  year = 2025,
  journal = {Journal of Computational Physics},
  volume = {525},
  issn = {00219991},
  chapter = {113732},
}

@article{xu_2025_prl,
	author = {Xu, F. and Liu, X.-S. and Li, X.-M. and Xia, K.-Q.},
	title = {Restoration of Axisymmetric Flow Structure in Turbulent Thermal Convection by Polymer Additives},
	journal = {Phys. Rev. Lett.},
	volume = {134},
	number = {8},
	ISSN = {0031-9007
	1079-7114},
	year = {2025},
	type = {Journal Article}
}

@article{zhang_pnas_2025_eecteea,
  title = {Experimental Evidence for the Continuous Transition between Elastic and Elastoinertial Turbulence in {Taylor-Couette} Flow},
  author = {Zhang, Y.-B. and Li, L. and Fan, Y.-N. and Su, J.-H. and Xi, H.-D. and Sun, C.},
  year = 2025,
  month = sep,
  journal = {Proc. Natl. Acad. Sci. U. S. A.},
  volume = {122},
  number = {38},
  pages = {e2505007122},
  issn = {0027-8424 1091-6490},
}

@article{zhou_prl_2002_psttma,
  title = {Plume Statistics in Thermal Turbulence: Mixing of an Active Scalar},
  author = {Zhou, S.-Q. and Xia, K.-Q.},
  year = 2002,
  journal = {Phys. Rev. Lett.},
  volume = {89},
  number = {18},
  pages = {184502--184502},
}

@article{Noether1918,
author = {Noether, E.},
journal = {Nachr. von Ges. Wiss. Gött. Math.-Phys. Kl.},
pages = {235-257},
title = {Invariante Variationsprobleme},
url = {http://eudml.org/doc/59024},
volume = {1918},
year = {1918},
}

@article{Chilla2012,
  title = {New perspectives in turbulent {{R}ayleigh-{B}\'enard} convection},
  author = {Chill{\`a}, F. and Schumacher, J.},
  year = 2012,
  journal = {Eur. Phys. J. E},
  volume = {35},
  pages = {58},
  issn = {12928941},
  doi = {10.1140/epje/i2012-12058-1}
}

@article{Gross1996,
  title = {The role of symmetry in fundamental physics},
  author = {Gross, David J.},
  year = 1996,
  journal = {Proc. Natl. Acad. Sci. U. S. A.},
  volume = {93},
  number = {25},
  pages = {14256--14259},
  issn = {0027-8424, 1091-6490},
  doi = {10.1073/pnas.93.25.14256},
  urldate = {2024-04-07}
}

@article{Krishnamurti1981,
  title = {Large-scale flow generation in turbulent convection},
  author = {Krishnamurti, Ruby and Howard, Louis N.},
  year = 1981,
  journal = {Proc. Natl. Acad. Sci. U. S. A.},
  volume = {78},
  number = {4},
  pages = {1981--1985},
  doi = {10.1073/pnas.78.4.1981}
}

@article{Sano1989,
  title = {Turbulence in helium-gas free convection},
  author = {Sano, Masaki and Wu, Xiao Zhong and Libchaber, Albert},
  year = 1989,
  journal = {Phys. Rev. A},
  volume = {40},
  number = {11},
  pages = {6421--6430},
  issn = {0556-2791},
  doi = {10.1103/PhysRevA.40.6421},
  urldate = {2024-08-20},
  copyright = {http://link.aps.org/licenses/aps-default-license}
}

@article{Verzicco1999,
  title = {Prandtl number effects in convective turbulence},
  author = {Verzicco, R. and Camussi, R.},
  year = 1999,
  journal = {J. Fluid Mech.},
  volume = {383},
  pages = {55--73},
  issn = {0022-1120, 1469-7645},
  doi = {10.1017/S0022112098003619},
  urldate = {2024-05-09},
  copyright = {https://www.cambridge.org/core/terms}
}

@book{Cantwell2002,
  title = {Introduction to symmetry analysis},
  author = {Cantwell, Brian J.},
  year = 2002,
  series = {Cambridge Texts Appl. Math.},
  publisher = {Cambridge University Press},
  isbn = {978-1-009-07445-2}
}

@article{Kadanoff2001,
  title = {Turbulent heat flow: Structures and scaling},
  author = {Kadanoff, Leo P.},
  year = 2001,
  journal = {Phys. Today},
  volume = {54},
  number = {8},
  pages = {34--39},
  issn = {00319228},
  doi = {10.1063/1.1404847}
}

@article{Gluckman1993,
  title = {Geometry of isothermal and isoconcentration surfaces in thermal turbulence},
  author = {Gluckman, B. J. and Willaime, H. and Gollub, J. P.},
  year = 1993,
  month = mar,
  journal = {Phys. Fluids A: Fluid Dyn.},
  volume = {5},
  number = {3},
  pages = {647--661},
  issn = {0899-8213},
  doi = {10.1063/1.858891},
  urldate = {2026-05-25}
}

@article{Peng1983,
  title = {Rheological behavior of {FM-9} solutions and correlation with flammability test results and interpretations},
  author = {Peng, S. T. J. and Landel, R. F.},
  year = 1983,
  journal = {J. Non-Newton. Fluid Mech.},
  volume = {12},
  number = {1},
  pages = {95--111},
  issn = {03770257},
  doi = {10.1016/0377-0257(83)80007-X},
  urldate = {2022-10-07}
}

@article{Tabor1986,
  title = {A cascade theory of drag reduction},
  author = {Tabor, M. and Gennes, P. G. De},
  year = 1986,
  journal = {Europhys. Lett.},
  volume = {2},
  number = {7},
  pages = {519--522},
  issn = {0295-5075, 1286-4854},
  doi = {10.1209/0295-5075/2/7/005},
  urldate = {2025-03-04}
}

@article{Benzi2018,
  title = {Polymers in fluid flows},
  author = {Benzi, Roberto and Ching, Emily S.C.},
  year = 2018,
  journal = {Annu. Rev. Condens. Matter Phys.},
  volume = {9},
  pages = {163--181},
  issn = {1947-5454, 1947-5462},
  doi = {10.1146/annurev-conmatphys-033117-053913},
  urldate = {2022-09-30}
}

@article{Herrchen1997,
  title = {A detailed comparison of various {FENE} dumbbell models},
  author = {Herrchen, Markus and {\"O}ttinger, Hans Christian},
  year = 1997,
  journal = {J. Non-Newton. Fluid Mech.},
  volume = {68},
  number = {1},
  pages = {17--42},
  issn = {03770257},
  doi = {10.1016/S0377-0257(96)01498-X},
  urldate = {2023-02-21}
}

@article{Serafini2022,
  title = {Drag reduction in turbulent wall-bounded flows of realistic polymer solutions},
  author = {Serafini, F. and Battista, F. and Gualtieri, P. and Casciola, C. M.},
  year = 2022,
  journal = {Phys. Rev. Lett.},
  volume = {129},
  number = {10},
  pages = {104502},
  issn = {0031-9007, 1079-7114},
  doi = {10.1103/PhysRevLett.129.104502},
  urldate = {2025-08-06}
}

@article{Xu2026,
  title = {Measured small-scale properties in turbulent {{R}ayleigh--{B}\'enard} convection with polymer additives},
  author = {Xu, Fang and Liu, Xiao-Shen and Li, Xiao-Ming and Xia, Ke-Qing},
  year = 2026,
  journal = {J. Fluid Mech.},
  volume = {1028},
  pages = {A31},
  doi = {10.1017/jfm.2026.11129}
}

@book{Bird1987a,
  title = {Dynamics of polymeric liquids},
  author = {Bird, Robert Byron and Armstrong, Robert C. and Hassager, Ole},
  year = 1987,
  edition = {2nd},
  volume = {Fluid Mech.},
  publisher = {Wiley},
  address = {New York},
  isbn = {978-0-471-80245-7}
}

@article{Perkins1995,
  title = {Stretching of a single tethered polymer in a uniform flow},
  author = {Perkins, Thomas T. and Smith, Douglas E. and Larson, Ronald G. and Chu, Steven},
  year = 1995,
  journal = {Science},
  volume = {268},
  number = {5207},
  pages = {83--87},
  issn = {0036-8075, 1095-9203},
  doi = {10.1126/science.7701345},
  urldate = {2026-01-14}
}

@article{Zhang1997a,
  title = {{Non-Boussinesq} effect: Thermal convection with broken symmetry},
  author = {Zhang, Jun and Childress, Stephen and Libchaber, Albert},
  year = 1997,
  journal = {Phys. Fluids},
  volume = {9},
  number = {4},
  pages = {1034--1042},
  issn = {1070-6631, 1089-7666},
  doi = {10.1063/1.869198},
  urldate = {2024-05-09}
}

@article{Ahlers2009b,
  title = {Heat transfer and large scale dynamics in turbulent {{R}ayleigh-{B}\'enard} convection},
  author = {Ahlers, Guenter and Grossmann, Siegfried and Lohse, Detlef},
  year = 2009,
  journal = {Reviews of Modern Physics},
  volume = {81},
  number = {2},
  pages = {503--537},
  issn = {00346861},
  doi = {10.1103/RevModPhys.81.503}
}

@article{Lohse2010,
  title = {Small-scale properties of turbulent {{R}ayleigh-{B}\'enard} convection},
  author = {Lohse, Detlef and Xia, Ke-Qing},
  year = 2010,
  journal = {Annu. Rev. Fluid Mech.},
  volume = {42},
  pages = {335--364},
  issn = {0066-4189},
  doi = {10.1146/annurev.fluid.010908.165152},
  langid = {american}
}

@article{Xia2013,
  title = {Current trends and future directions in turbulent thermal convection},
  author = {Xia, Ke-Qing},
  year = 2013,
  journal = {Theor. Appl. Mech. Lett.},
  volume = {3},
  number = {5},
  pages = {052001},
  publisher = {Elsevier},
  issn = {20950349},
  doi = {10.1063/2.1305201}
}

@book{Schwichtenberg2018,
  title = {Physics from symmetry},
  author = {Schwichtenberg, Jakob},
  year = 2018,
  series = {Undergraduate Lecture Notes in Physics},
  publisher = {Springer International Publishing},
  address = {Cham},
  doi = {10.1007/978-3-319-66631-0},
  urldate = {2024-04-22},
  copyright = {http://www.springer.com/tdm},
  isbn = {978-3-319-66630-3 978-3-319-66631-0}
}

@article{Williams2025,
  title = {Asymmetries in nominally symmetric flows},
  author = {Williams, Owen J. H. and Smits, Alexander J.},
  year = 2025,
  journal = {Annu. Rev. Fluid Mech.},
  volume = {57},
  pages = {35--60},
  issn = {0066-4189, 1545-4479},
  doi = {10.1146/annurev-fluid-030124-045719},
  urldate = {2024-09-04}
}

@article{Shang2004,
  title = {Measurements of the local convective heat flux in turbulent {{R}ayleigh-{B}\'enard} convection},
  author = {Shang, X.-D. and Qiu, X.-L. and Tong, P. and Xia, K.-Q.},
  year = 2004,
  journal = {Phys. Rev. E},
  volume = {70},
  number = {2},
  pages = {026308},
  issn = {1063651X},
  doi = {10.1103/PhysRevE.70.026308},
  langid = {american}
}

@article{Datta2022,
  title = {Perspectives on viscoelastic flow instabilities and elastic turbulence},
  author = {Datta, Sujit S. and Ardekani, Arezoo M. and Arratia, Paulo E. and Beris, Antony N. and Bischofberger, Irmgard and McKinley, Gareth H. and Eggers, Jens G. and {L{\'o}pez-Aguilar}, J. Esteban and Fielding, Suzanne M. and Frishman, Anna and Graham, Michael D. and Guasto, Jeffrey S. and Haward, Simon J. and Shen, Amy Q. and Hormozi, Sarah and Morozov, Alexander and Poole, Robert J. and Shankar, V. and Shaqfeh, Eric S. G. and Stark, Holger and Steinberg, Victor and Subramanian, Ganesh and Stone, Howard A.},
  year = 2022,
  journal = {Phys. Rev. Fluids},
  volume = {7},
  number = {8},
  pages = {080701},
  issn = {2469-990X},
  doi = {10.1103/PhysRevFluids.7.080701},
  urldate = {2023-02-16}
}

@article{Xu2024,
  title = {Experimental measurement of spatio-temporally resolved energy dissipation rate in turbulent {{R}ayleigh--{B}\'enard} convection},
  author = {Xu, Fang and Zhang, Lu and Xia, Ke-Qing},
  year = 2024,
  journal = {J. Fluid Mech.},
  volume = {984},
  pages = {A8},
  doi = {10.1017/jfm.2024.164}
}

@article{Xie2015,
  title = {Effects of polymer additives in the bulk of turbulent thermal convection},
  author = {Xie, Yi-Chao and Huang, Shi-Di and Funfschilling, Denis and Li, Xiao-Ming and Ni, Rui and Xia, Ke-Qing},
  year = 2015,
  journal = {J. Fluid Mech.},
  volume = {784},
  pages = {R3},
  issn = {14697645},
  doi = {10.1017/jfm.2015.618}
}

@article{Moisy2001,
  title = {Passive scalar intermittency in low temperature Helium flows},
  author = {Moisy, F. and Willaime, H. and Andersen, J. S. and Tabeling, P.},
  year = 2001,
  journal = {Phys. Rev. Lett.},
  volume = {86},
  number = {21},
  pages = {4827--4830},
  issn = {0031-9007, 1079-7114},
  doi = {10.1103/PhysRevLett.86.4827},
  urldate = {2023-04-04}
}

@article{Xi2004,
  title = {From laminar plumes to organized flows: the onset of large-scale circulation in turbulent thermal convection},
  author = {Xi, Heng-Dong and Lam, Siu and Xia, Ke-Qing},
  year = 2004,
  journal = {J. Fluid Mech.},
  volume = {503},
  pages = {47--56},
  issn = {00221120},
  doi = {10.1017/S0022112004008079}
}

@article{Benzi2010,
  title = {Effect of polymer additives on heat transport in turbulent thermal convection},
  author = {Benzi, Roberto and Ching, Emily S. C. and De Angelis, Elisabetta},
  year = 2010,
  journal = {Phys. Rev. Lett.},
  volume = {104},
  number = {2},
  pages = {024502},
  issn = {00319007},
  doi = {10.1103/PhysRevLett.104.024502}
}

\end{document}